\documentclass[aps,pra,amssymb,amsmath,twocolumn,float]{revtex4-2}
\usepackage{graphicx}
\usepackage{amsmath, mathtools}
\usepackage{xcolor}
\usepackage{float}

\mathtoolsset{showonlyrefs=true}

\begin{document}
\title{Determining the boundary of dynamical chaos in the generalized Chirikov map via machine learning}
\author{Daniil Chernyshov}
\author{Arkady Satanin}
\author{Lev Shchur}
\affiliation{Computational Physics Laboratory, HSE University, 101000 Moscow, Russia}

\begin{abstract}
We investigate the boundary separating regular and chaotic dynamics in the generalized Chirikov map, an extension of the standard map with a control parameter $K$ and a phase-shifted (phase $\tau$) secondary sequence of kicks with a control parameter $K_\alpha$. Lyapunov maps were computed across the parameter space $(K, K_\alpha; \tau)$ and used to train a convolutional neural network (ResNet18) for binary classification of dynamical regimes. The trained model reproduces the known critical control parameter $K_c$ for the onset of global chaos in the standard map and identifies two-dimensional boundaries $(K, K_\alpha)$ in the generalized map for varying phase shifts $\tau$. The results reveal systematic deformation of the boundary as $\tau$ increases, highlighting the sensitivity of the system to phase modulation and demonstrating the ability of machine learning to extract interpretable features of complex Hamiltonian dynamics. This framework allows precise characterization of stability boundaries in nontrivial nonlinear systems.
\end{abstract}

\maketitle

\textit{Introduction. ---} Many fundamental problems in physics reduce to the study of Hamiltonian systems with several degrees of freedom. Examples include the restricted three-body problem, the motion of atomic chains in a periodic potential (the Frenkel–Kontorova model), the description of particle orbits in accelerators, and the acceleration of cosmic particles (Fermi acceleration)~\cite{Lichtenberg-2013}. At the same time, the equations governing the dynamics of nonlinear systems are, in the most interesting cases, typically non-integrable. According to Poincaré, non-integrability arises from the presence of unstable hyperbolic points in phase space, which gives rise to homoclinic structures of trajectories near the separatrix~\cite{Poincare-1967, Kozlov-1983}. In dynamical systems, variations of parameters may lead to resonance capture (both primary and secondary), the formation of stochastic layers around the separatrix, resonance overlap, and ultimately the onset of global chaos~\cite{Chirikov-1971, Chirikov-1979} with the fractal boundaries between the regular and chaotic regions~\cite{Zaslavsky-2005}. The scenario outlined above is rather general; however, for practically relevant systems, fundamental questions remain unanswered: \textit{Does a boundary exist between chaos and order in phase space?} \textit{What is the structure of the fractal geometry of stochastic regions?}

For periodically driven systems, corresponding to Hamiltonian systems with $3/2$ degrees of freedom, the qualitative scenario of chaos development, following the description of Chirikov~\cite{Chirikov-1971, Chirikov-1979}, can be outlined as follows. As the amplitude of the external periodic force increases, or equivalently as the control parameter $K$ in the standard map increases, the system first undergoes resonance capture. Owing to periodicity, chains of resonances then appear, around which closed invariant curves are formed. A further increase of the perturbation results in the separatrix becoming embedded in a stochastic layer, eventually leading to resonance overlap. As the amplitude increases, a \textit{stochastic sea} forms in phase space (in the Poincaré map), within which small islands of regular motion are embedded. Chirikov’s resonance-overlap scenario provides insight into the mechanism underlying the formation of the global structure of chaos; however, the arguments presented are not rigorous and are largely based on numerical experiments. The most accurate value of the control parameter for the Chirikov map was determined in~\cite{Greene-1979}. Further accounts of recent advances in the theory of dynamical chaos in two-dimensional systems can be found in~\cite{Zaslavsky-2005, Reichl-2021, Mugnaine-2024} and the references therein.

In this work, we focus on the properties of the boundary between regular and stochastic motion, using the standard map — the Chirikov map — as an example. It is well known that in the case of the Chirikov map, resonance overlap appears only at the second order of perturbation theory (the overlap of \textit{virtual resonances}). We consider a generalized Chirikov map, which represents a natural extension of the standard map through the introduction of phase-shift parameter and an additional external forcing ("binary signal"). This modification leads to a more complex structure of phase-space regions already at the leading order, allowing the inclusion of additional control mechanisms for chaos management and opening new possibilities for identifying the boundary between regular and chaotic dynamics in nontrivial cases. As noted by Chirikov~\cite{Chirikov-1979}, even a single trajectory of a nonlinear dynamical system carries a wealth of information about the system, making the processing of numerical data a significant challenge. Importantly, in the Chirikov map, the system trajectory between successive perturbations is known exactly, ensuring numerical stability of the iteration procedure. The map considered in this work preserves this property, allowing the generation of long trajectories. 

Recent progress in addressing various problems in the theory of dynamical systems has been driven by advances in modern supercomputing and quantum computation technologies, as well as the application of neuromorphic methods for data processing and analysis. As an example of using recognition, classification, and large-scale data analysis methods in the study of the complex systems and critical phenomena, see~\cite{Carrasquilla-2017, Chertenkov-2023, Lee-2020, Celletti-2022, Barrio-2023, Zu-2025}. In the present work, we propose to employ supervised machine learning techniques to determine the dynamical chaos boundary of a system, i.e., to identify the boundary between chaos and order.

To determine the boundary separating regular and chaotic dynamical regimes in the generalized Chirikov map, modern methods of data analysis using supervised machine learning will be employed. First, training and validation datasets will be generated in the form of Lyapunov maps corresponding to various parameters of the standard map. Next, a suitable neural network architecture will be selected. This will be followed by data preparation for training the neural network and the network training itself. Based on these calculations, a binary classification of Lyapunov maps of the generalized Chirikov map will be performed using the trained neural network.

Based on the conducted analysis, and also drawing on Greene’s theory~\cite{Greene-1979} in the extended parameter space, the boundary between chaos and regular dynamics will be described.

\textit{Generalized Chirikov map. ---} 
\begin{figure}
\centering
\includegraphics[width=0.45\textwidth]{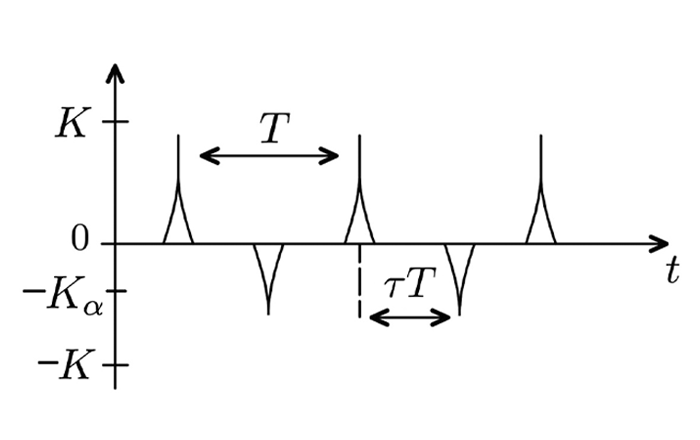}
\caption{Binary sequence of kicks. Vertical axis: dimensionless amplitude of the primary kicks, \(K\), applied at times \(nT\); \(K_\alpha\) denotes the amplitude of the secondary kicks applied at times \(nT+\tau T\) (phase shift \(\tau T\)).}
\label{fig:fig1}
\end{figure}
The Hamiltonian of a planar rotator subjected to a binary sequence of kicks, schematically illustrated in Figure~\ref{fig:fig1}, can be written as:

\begin{multline}
\label{eq:bin_signal}
H = \frac{I^2}{2} + \cos \theta \Biggl[
K \sum_{n=-\infty}^{\infty} \delta(t - nT) \\
- K_\alpha \sum_{n=-\infty}^{\infty} \delta(t - T(n+\tau))
\Biggr],
\end{multline}

\noindent
where $(I, \theta)$ denote the generalized momentum of the rotator with unit moment of inertia and the generalized coordinate (angle), respectively; $K$ and $K_\alpha$ are the constants characterizing the action of the external force on the rotator; $T$ is the period of the external force; and $\tau T$ is the phase shift.

Since no external force acts on the rotator between kicks, the generalized momentum remains constant, while the phase evolves with a uniform angular velocity. This allows one to directly write down the generalized map for the corresponding pairs of variables. Let $I_n \equiv I(nT)$ and $\theta_n \equiv \theta(nT)$ be defined immediately before the kick applied at time $nT$. Under the action of the "positive" kick at $nT$, the momentum changes according to $I_{n+\tau} = I_n + K \sin \theta_n$, after which the phase evolves freely and, over the time interval $\tau T$, acquires an increment of $\tau I_{n+\tau}$ (with $I_{n+\tau} \equiv I(nT + \tau T)$ and $\theta_{n+\tau} \equiv \theta(nT + \tau T)$). A similar update of the action variable and phase occurs under the "negative" kick at $(n + \tau)T$. Consequently, after one period, the combined map takes the form:
no

It should be noted that we have rescaled the variables as follows: 
$I_n T \rightarrow I_n, \quad K T \rightarrow K, \quad K_\alpha T \rightarrow K_\alpha$.

By eliminating the intermediate variables $I_{n+\tau}$ and $\theta_{n+\tau}$, we obtain:
\begin{equation}
\label{eq:simple_gcm}
\begin{minipage}{0.9\columnwidth}
\[
\left\{
\begin{aligned}
I_{n+1} &= I_n + K \sin \theta_n \\
&\quad - K_\alpha \sin \!\bigl(\theta_n + \tau(I_n + K \sin \theta_n)\bigr), \\
\theta_{n+1} &= \theta_n + \tau (I_n + K \sin \theta_n) + (1 - \tau) I_{n+1}.
\end{aligned}
\right.
\]
\end{minipage}
\end{equation}

It is evident that for $K_\alpha = 0$, the map~\eqref{eq:simple_gcm} reduces to the Chirikov map, while in the case $K = 0$ and $K_\alpha \neq 0$, the map again (after the phase shift) reduces to the Chirikov map. 

Note that, due to the periodicity of the $\sin \theta$ function, the system dynamics can be conveniently considered on a torus by taking $[(I, \theta) \bmod 2\pi]$. Mathematically, the standard map is a two-dimensional Hamiltonian (symplectic) map that preserves phase-space volume, i.e., its Jacobian is equal to 1 at every point in the phase space.

Let us first discuss the case $K_\alpha = 0$. As the coupling constant $K$ increases, the resonances overlap and are destroyed according to the scenario developed by Chirikov~\cite{Chirikov-1971, Chirikov-1979}, after which secondary resonances emerge within the stochastic sea. The value of $K$ at which the invariant torus is destroyed and global chaos appears has been determined by various authors. Greene's method is the most precise of those developed to date, yielding $K_c \approx 0.971635406$~\cite{Greene-1979}.

To further quantify the system's dynamics and the onset of chaos, we analyze the Lyapunov exponents, which measure the rate of exponential divergence of trajectories in phase space. A trajectory on the torus is generated by choosing an initial point with coordinates $(I_0, \theta_0)$. Iterating the map~\eqref{eq:simple_gcm} $N$ times yields the trajectory $(I_0, \theta_0), \dots, (I_n, \theta_n), \dots, (I_N, \theta_N)$. If the initial conditions are slightly perturbed, the trajectory changes to $(I_n + \delta I_n, \theta_n + \delta \theta_n)$. The maximal Lyapunov exponent is defined as $\Lambda = \lim\limits_{N \to \infty} \Lambda_N$, where

\begin{equation}
\Lambda_N = \frac{1}{N} \sum_{n=1}^{N} \ln \frac{d_n}{d_{n-1}}.
\end{equation}

\noindent
where $d_n = \sqrt{\delta I_n^2 + \delta \theta_n^2}$, with $d_0$ being the initial distance between nearby points in phase space. The infinitesimal deviations are obtained by jointly solving~\eqref{eq:simple_gcm} and

\begin{equation}
\label{eq:joint}
\begin{minipage}{0.9\columnwidth}
\[
\left\{
\begin{aligned}
\delta I_{n+1} &= (1 - \tau K_\alpha \cos(\theta_n + \tau (I_n + K \sin\theta_n))) \, \delta I_n \\
&\quad + \Bigl[ K \cos\theta_n 
       - K_\alpha \cos(\theta_n + \tau (I_n + K \sin\theta_n)) \\
&\qquad \times (1 + \tau K \cos\theta_n) \Bigr] \delta \theta_n, \\[1ex]
\delta \theta_{n+1} &= (1 + \tau K \cos\theta_n) \, \delta \theta_n \\
&\quad + \tau \delta I_n + (1 - \tau) \delta I_{n+1}.
\end{aligned}
\right.
\]
\end{minipage}
\end{equation}

The system~\eqref{eq:joint} allows us to investigate the stability of fixed points of the map~\eqref{eq:simple_gcm}. For example, in the case $K_\alpha = 0$, the stationary point $(0, \pi)$ is found to be a stable elliptic point for $0 \leq K \leq 4$. For $K \geq 4$, the stationary point $(0, \pi)$ loses stability and becomes hyperbolic. Similarly, the fixed point $(0, \pi)$, which is elliptic when $K_\alpha = 0$, turns into an unstable hyperbolic point for $K = 0$ and $K_\alpha \neq 0$. Interestingly, when $K = K_\alpha$, the map~\eqref{eq:simple_gcm} exhibits fixed points that depend on the phase $\tau$, while the point $(0, \pi)$ remains stable for any value of $\tau$.

In conclusion of this section, we note that a detailed study of the fine structure of the phase portrait generated by the Chirikov map, which exhibits fractal properties, remains an open problem to date.

\textit{Data generation. ---} The generalized Chirikov map~\eqref{eq:simple_gcm} exhibits both regular and chaotic dynamical regimes depending on the parameters $(K, K_\alpha; \tau)$. To classify the dynamical regimes based on the maximal Lyapunov exponent in the phase space $(I, \theta) \in [0, 2\pi]^2$, supervised machine learning methods are employed. This approach allows formalizing the problem of identifying the nature of a trajectory as a binary image classification task~\cite{Goodfellow-2016, Krizhevsky-2012}.

For training and validation, datasets of Lyapunov maps were generated for the standard map $(K_\alpha \equiv 0)$ at three resolutions ($64 \times 64$, $128 \times 128$, and $256 \times 256$ pixels) to compare the effect of input size on network performance. Each image corresponds to a phase portrait in which the horizontal axis represents the variable $\theta \in [0, 2\pi]$, the vertical axis represents $I \in [0, 2\pi]$, and the color of each point encodes the value of the maximal Lyapunov exponent at the corresponding location in phase space: "warmer" colors indicate larger exponent values, whereas "cooler" colors correspond to smaller ones.

\begin{figure}
    \centering
    \includegraphics[width=0.48\textwidth]{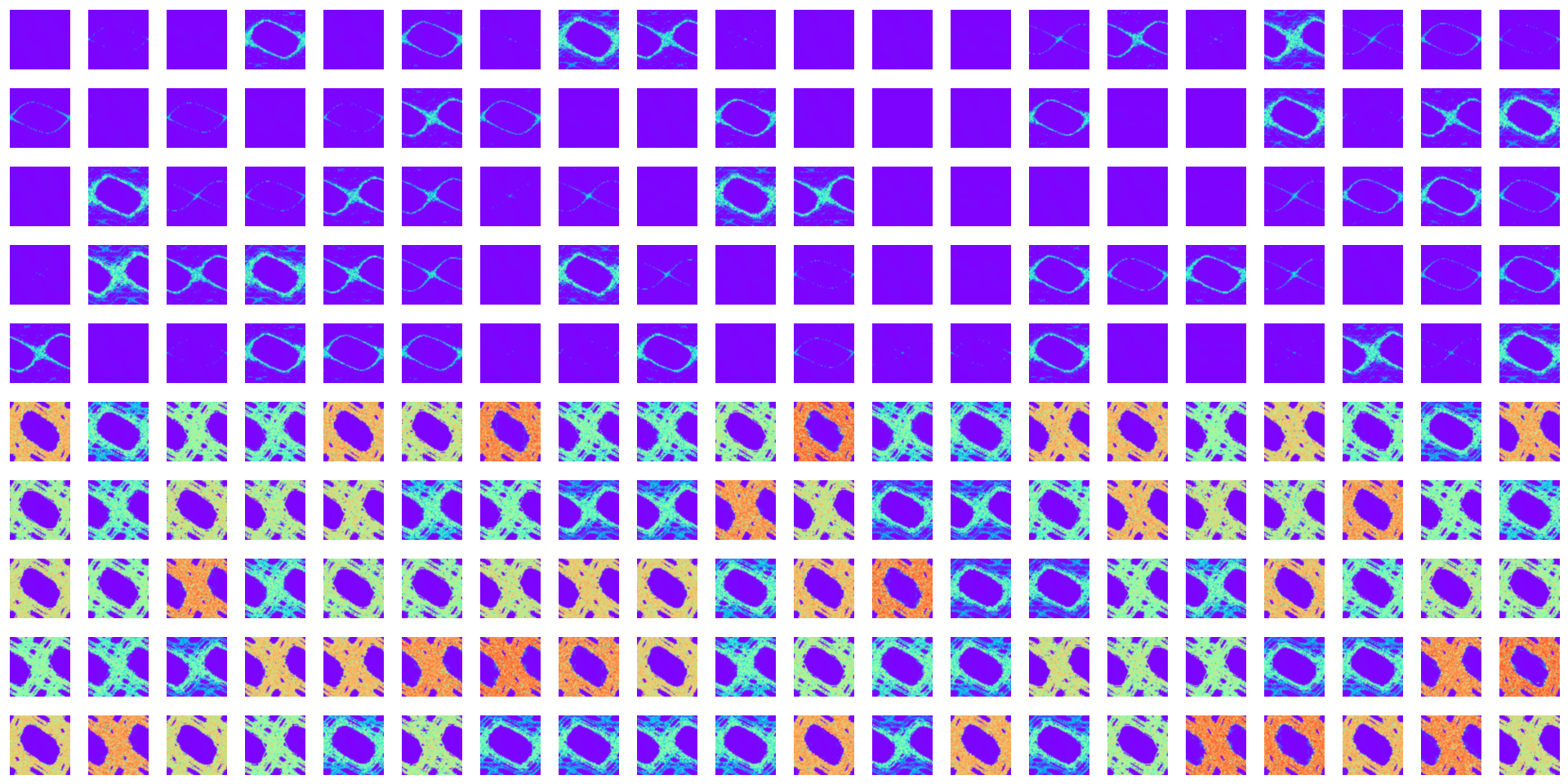}
    \caption{Example of Lyapunov maps ($64 \times 64$ pixels) used for training and validation of the neural network. The top five rows correspond to regular dynamics, while the bottom rows correspond to chaotic dynamics.}
    \label{fig:fig2}
\end{figure}

Using numerical simulations of the standard map and subsequent computation of Lyapunov exponents~\cite{Benettin-1980}, three separate datasets were generated, each containing 400 Lyapunov maps at different resolutions. Each dataset comprises 200 maps corresponding to regular dynamics ($K < K_c$) and 200 maps corresponding to chaotic dynamics ($K > K_c$). To enhance symmetry learning in the model, within each regime 100 images were generated using the standard map ($K \neq 0$, $K_\alpha = 0$) and 100 images using the generalized Chirikov map with parameters equivalent to those of the standard map ($K = 0$, $K_\alpha \neq 0$).

Within each regime, the control parameters were sampled uniformly. For the standard map, $K \sim \mathcal{U}(0, K_c - 1 \times 10^{-4})$ for regular dynamics and $K \sim \mathcal{U}(K_c, 1.5)$ for chaotic dynamics; for the generalized Chirikov map, $K_\alpha$ was sampled analogously, $K_\alpha \sim \mathcal{U}(0, K_c - 1 \times 10^{-4})$ for regular dynamics and $K_\alpha \sim \mathcal{U}(K_c, 1.5)$ for chaotic dynamics. Each image was assigned a binary label: 0 for regular dynamics and 1 for chaotic dynamics. Examples of the generated $64 \times 64$ images are shown in Figure~\ref{fig:fig2}.

\begin{figure}
\centering
\includegraphics[width=0.5\textwidth]{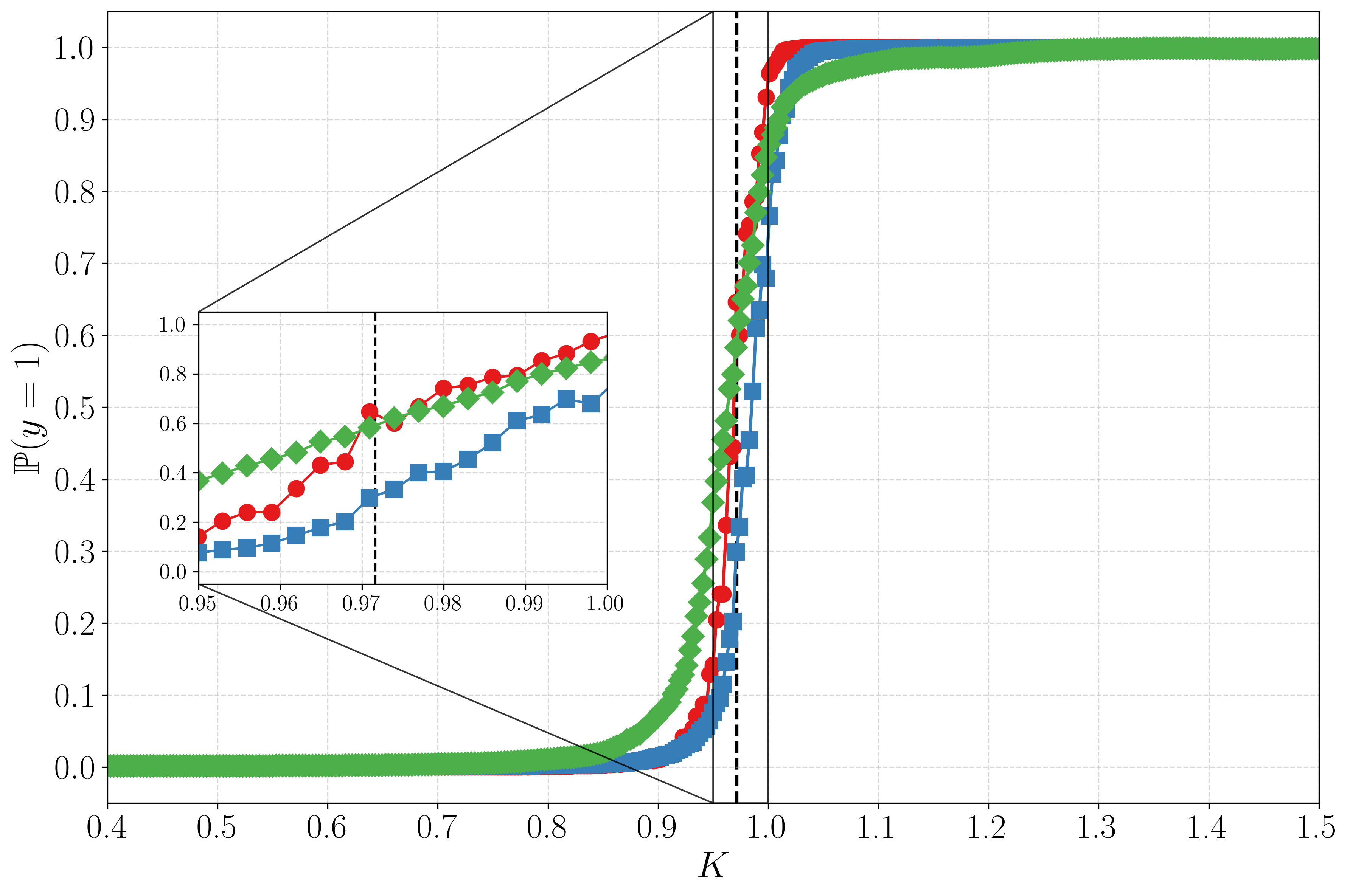}
\caption{Probability of classifying the Lyapunov map as chaotic, $\mathbb{P}(y = 1)$, predicted by the neural network for different input image resolutions. Curves correspond to resolutions $64 \times 64$ (red circles), $128 \times 128$ (blue squares), and $256 \times 256$ (green diamonds). The vertical dashed line indicates the critical value $K_c$. An inset presents a magnified view of the transition region in the vicinity of $K_c$.}
\label{fig:fig3}
\end{figure}

\textit{Neural network model. ---}
To address the binary classification task of Lyapunov maps, three convolutional neural networks~\cite{OShea-2015} based on ResNet18~\cite{He-2016} were trained, each corresponding to a different input resolution ($64 \times 64$, $128 \times 128$, and $256 \times 256$ pixels). For each network, the first convolutional layer was adapted to the input size, and the initial max-pooling operation was removed to preserve feature resolution in the early layers. The final fully connected layer was replaced by a two-output layer corresponding to the classes \textit{regular} and \textit{chaotic} dynamics. Subsequent analysis shows that higher-resolution images ($256 \times 256$) yield improved classification performance compared to lower-resolution inputs (Figure~\ref{fig:fig3}).

The choice of the ResNet architecture is motivated by its well-established connection to the theory of dynamical systems and numerical methods for differential equations. In~\cite{Ruthotto-2020}, it was shown that residual blocks can be interpreted as a time discretization of evolutionary partial differential equations, providing a rigorous mathematical connection between the network depth and the integration steps of the dynamical system. In~\cite{Chashchin-2019}, it was demonstrated that ResNet is capable of reproducing trajectories of ordinary differential equations and predicting their evolution stably over horizons much longer than the training data. Furthermore,~\cite{Liu-2024} proposed a multilevel approach to modeling spatiotemporal dynamics, where ResNet blocks are used to preserve the structure of the corresponding differential equations, enabling efficient representation of complex spatiotemporal processes. These findings indicate that the ResNet architecture not only achieves high performance in traditional computer vision tasks but also has significant potential for modeling and analyzing complex dynamical systems.

\textit{Data preprocessing and training. ---} Prior to training, standard data augmentation techniques were applied to the Lyapunov map images to improve the generalization capability of the neural network. Specifically, random horizontal and vertical flips, as well as rotations up to $20^\circ$, were performed~\cite{Shorten-2019}.

Following these augmentations, the full dataset of 400 images was uniformly shuffled to preserve class balance and then split into training and validation sets with an 80:20 ratio. The model was trained using the Adam optimizer~\cite{Kingma-2014} with parameters $\alpha = 10^{-3}$, $\beta_1 = 0.9$, $\beta_2 = 0.999$, $\varepsilon = 10^{-8}$, and weight decay $\lambda = 10^{-4}$, in combination with a cosine annealing learning rate schedule with $T_{\max} = 10$. Binary cross-entropy was used as the loss function, and the network was trained for 20 epochs. The output probability $\mathbb{P}(y=1 \mid \mathbf{x})$ that the input Lyapunov map $\mathbf{x}$ corresponds to chaotic dynamics ($y=1$) is defined as $\mathbb{P}(y=1 \mid \mathbf{x}) = 1/(1+e^{-f(\mathbf{x})})$, where $f(\mathbf{x})$ is the logit (pre-activation value) produced by the final layer of the network.

\begin{figure}
\centering
\includegraphics[width=0.5\textwidth]{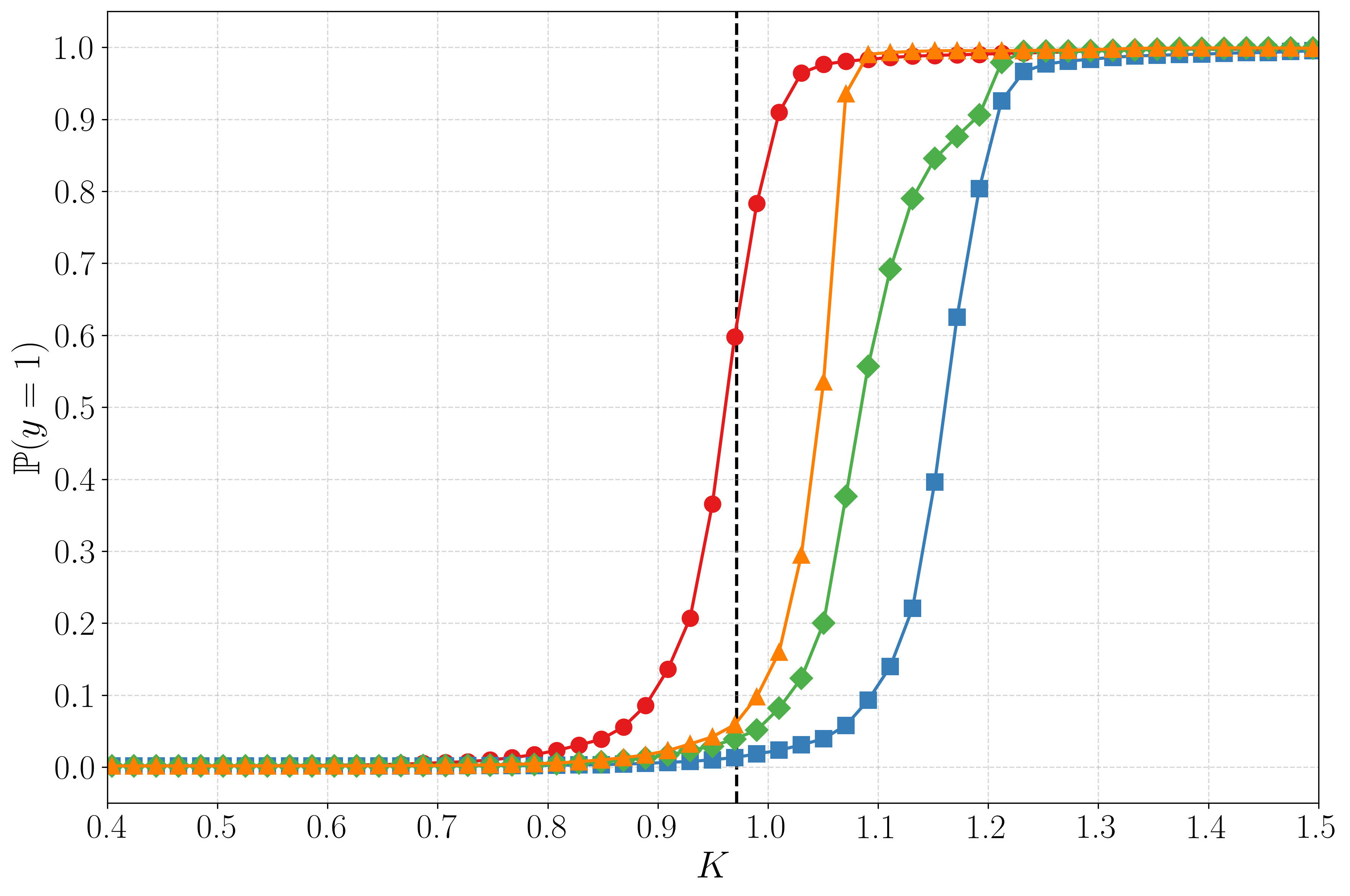}
\caption{Probability of classifying the Lyapunov map as chaotic, $\mathbb{P}(y = 1)$, predicted by the neural network as a function of the parameter $K$ in the generalized Chirikov map. Curves correspond to the following parameter sets: ($K_\alpha=0$, $\tau=0$) (red circles); ($K_\alpha=0.2$, $\tau=0.5$) (orange upward triangles); ($K_\alpha=0.2$, $\tau=0.25$) (green diamonds);  and ($K_\alpha=0.2$, $\tau=0$) (blue squares). The vertical dashed line indicates the critical value $K_c$.}
\label{fig:fig4}
\end{figure}

\textit{Results analysis. ---} 
Further analysis, performed using the network trained on the highest-resolution images ($256 \times 256$ pixels), allows us to assess the sensitivity of the trained model to variations in the parameter $K$. Figure~\ref{fig:fig4} shows the dependence of the probability of classifying a Lyapunov map as chaotic on the value of the parameter $K$ in the generalized Chirikov map.

The obtained results are in good agreement with the established theory. For $K_\alpha = 0$ and $\tau = 0$ (red circles), the transition from regular to chaotic dynamics occurs near the critical value $K_c \approx 0.97$. Following Greene’s argument~\cite{Greene-1979}, this threshold for the Chirikov map is defined by the value of the coupling parameter $K$ at which the last invariant curve on the two-dimensional torus is destroyed.
In the generalized Chirikov map, two additional parameters, $K_\alpha$ and $\tau$, influence the dynamics. When $\tau = 0$, the dependence of the critical boundary on $K_\alpha$ (Figure~\ref{fig:fig4}) can be attributed to the effective reduction of the coupling constant in the Hamiltonian~\eqref{eq:bin_signal}, which delays the destruction of the invariant curve. When $\tau \ne 0$, the introduction of a second series of $\delta$-functions explicitly depends on $\tau$, making the shape of the critical boundary phase-dependent. For instance, for $K_\alpha = 0.2$ and $\tau = 0$ (blue squares), the transition boundary shifts to the right along the $K$-axis, as the additional kicks compensate the primary ones by the amount $K_\alpha$.

It is of particular interest to examine the response of the neural network to the introduction of the parameter $\tau$, which modulates the phase of the additional sequence of kicks in~\eqref{eq:simple_gcm}.

\begin{figure*}[t]
\centering
\includegraphics[width=\textwidth]{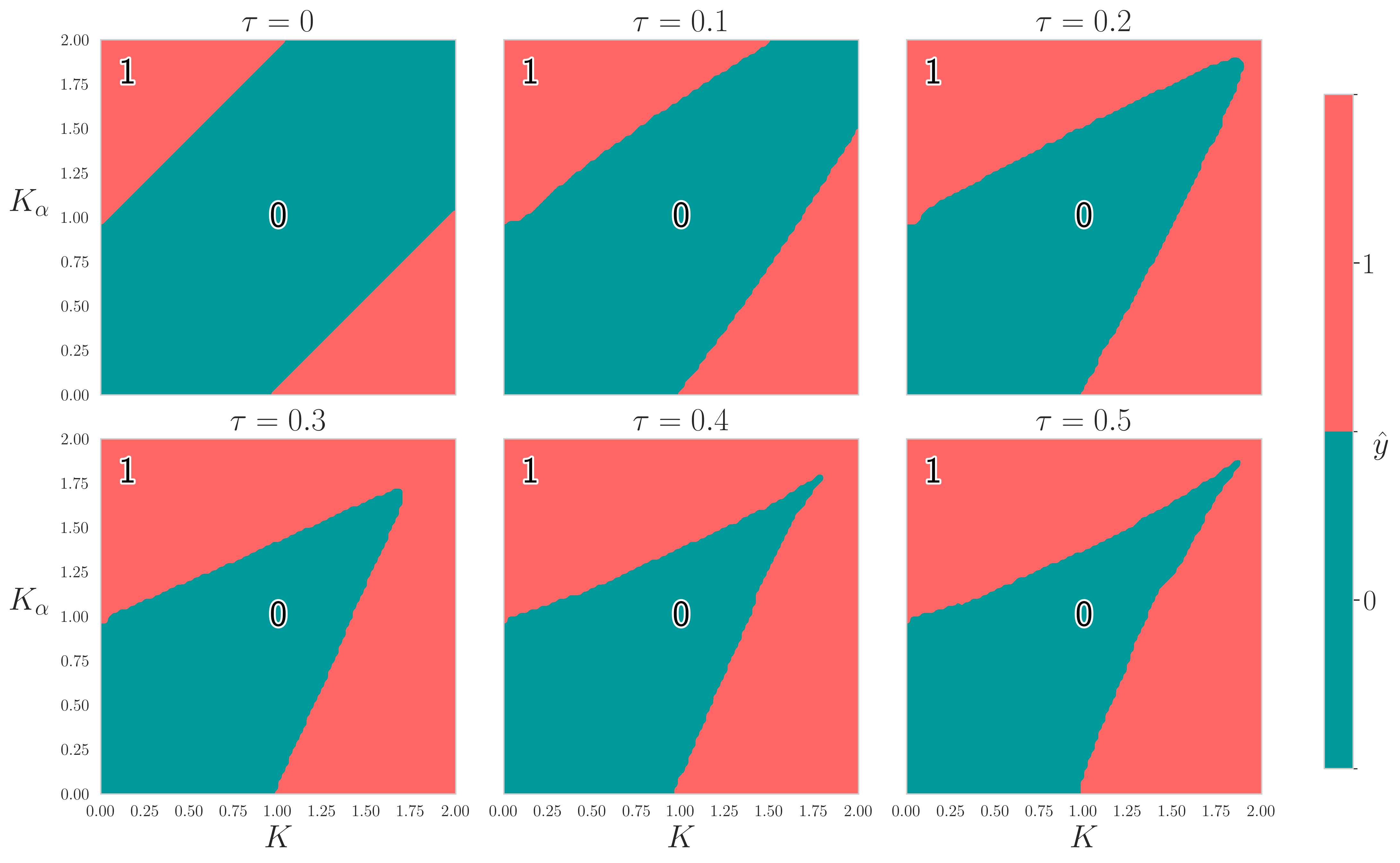}
\caption{Neural network predictions $\hat{y}$ on a uniform $100 \times 100$ grid in the $(K, K_\alpha)$ parameter plane for $\tau = 0, 0.1, 0.2, 0.3, 0.4, 0.5$. Regular regime: $\hat{y}=0$; chaotic regime: $\hat{y}=1$.}
\label{fig:fig5}
\end{figure*}

Figure~\ref{fig:fig5} shows the predictions $\hat{y}$ obtained on a uniform grid in the $(K, K_\alpha)$ parameter plane for different values of $\tau$. These results were obtained by computing Lyapunov maps of the generalized Chirikov map, which were then provided as input to the neural network to classify the corresponding dynamical regime. Predictions for cases with $\tau > 0$ were obtained in the same manner.

The model demonstrates a robust ability to distinguish between regular and chaotic dynamical regimes. In particular, the predictions exhibit a clear symmetry with respect to the parameter exchange $(K, K_\alpha) \mapsto (K_\alpha, K)$, which reflects the partial compensation of the $\delta$-kicks in~\eqref{eq:simple_gcm}.

As $\tau$ increases, this symmetry gradually deteriorates: the upper corners of the chaotic domains approach one another, forming a wedge-like structure. While for $\tau = 0$ the boundary between regular and chaotic regions is approximately linear and can be expressed as $K_\alpha \approx aK + b$, for $\tau > 0$ it takes the more general form $K_\alpha \approx f(K, \tau)$, where $f$ denotes a numerically observed dependence capturing the effect of the phase shift. This observation highlights the sensitivity of the system to the phase parameter $\tau$ while preserving the interpretability of the model’s predictions.

\textit{Discussion. ---}
In summary, the proposed methodology for data generation and analysis in the context of the generalized Chirikov map enables the identification of fine features of the boundary separating regular and chaotic dynamical regimes using machine learning techniques. The analysis demonstrates that the neural network successfully reproduces known results regarding the critical parameter for the onset of global chaos in the standard map.

For the generalized Chirikov map, the network exhibits satisfactory performance in distinguishing between regular and chaotic dynamical regimes, including the presence of two-dimensional boundaries separating these regimes, boundary curve structures consistent with Greene's theory~\cite{Greene-1979}, and the emergence of phase-dependent structures in mixed regimes.

The results indicate that increasing the dimensionality of the parameter space leads to a more complex structure of the boundary of dynamical chaos.

\begin{acknowledgments}
This research was supported in part through computational resources of HPC facilities at HSE University.

This article is an output of a research project (HSE-BR-2025-007) implemented as part of the Basic Research Program at HSE University.
\end{acknowledgments}

\section*{Author Contributions}
D.C. and A.S. contributed equally to this work. L.S. conceptualized the work.

\end{document}